\documentclass[prl,showpacs,floatfix,twocolumn,byrevtex]{revtex4-1}
\pdfoutput=1
\usepackage{amsmath}
\usepackage{amssymb}
\usepackage{amstext}
\usepackage{amsopn}
\usepackage{amsfonts}
\usepackage{amsxtra}
\usepackage{color}
\usepackage{dcolumn}
\usepackage{graphicx}
\usepackage{hyperref}
\usepackage{bm}
\begin{document}

\preprint{Draft --- not for distribution}

%
%
\title{Spectral weight transfer in strongly-correlated \boldmath Fe$_{1.03}$Te \unboldmath}
\author{Y. M. Dai}
\author{A. Akrap}
\altaffiliation{Present address: University of Geneva, CH-1211 Geneva 4, Switzerland}
\author{J. Schneeloch}
\author{R. D. Zhong}
\author{T. S. Liu}
\altaffiliation{School of Chemical Engineering and Environment, North University of China,
  Taiyuan, Shanxi PRC 030051}
\author{G. D. Gu}
\author{Q. Li}
\author{C. C. Homes}
\email{homes@bnl.gov}
\affiliation{Condensed Matter Physics and Materials Science Department,
  Brookhaven National Laboratory, Upton, New York 11973, USA}%
\date{\today}

%
%
\begin{abstract}
The temperature dependence of the in-plane optical conductivity has been determined for
Fe$_{1.03}$Te above and below the magnetic and structural transition at $T_N\simeq 68$~K.
The electron and hole pockets are treated as two separate electronic subsystems;
a strong, broad Drude response that is largely temperature independent, and a much
weaker, narrow Drude response with a strong temperature dependence. Spectral weight
is transferred from high to low frequency below $T_N$, resulting in the dramatic
increase of both the low-frequency conductivity and the related plasma frequency.
The change in the plasma frequency is due to an increase in the carrier concentration
resulting from the closing of the pseudogap on the electron pocket, as well as the likely
decrease of the effective mass in the antiferromagnetic state.
\end{abstract}
%
%
%
%
%
%
%
%
%
%
\maketitle

%
%
%
The discovery of superconductivity at high temperatures in the
iron-arsenic materials \cite{kamihara08,ren08,johnston10} prompted the search for
this phenomenon in other iron-based systems, including the iron-chalocogenide
materials. FeSe is an ambient-pressure superconductor with a critical
temperature $T_c \simeq 8$~K; while the partial substitution of Te for Se almost
doubles the $T_c$, the end compound FeTe is not a superconductor \cite{khasanov09}.
For nearly stoichiometric Fe$_{1+\delta}$Te there is a first-order structural and magnetic
transition \cite{bao09,zaliznyak11,zaliznyak12} from a tetragonal, paramagnetic (PM) state to a
monoclinic, antiferromagnetic (AFM) state at $T_N \simeq 68$~K.  In the low-temperature phase
a bond-order wave is observed that has been attributed to ferro-orbital ordering \cite{fobes14}.
At ambient pressure, the structural and magnetic properties depend strongly on the amount
of excess Fe \cite{liu11,rodriguez11}.  Fe$_{1+\delta}$Te is a poor metal that exhibits a
non-metallic resistivity that increases with decreasing temperature \cite{chen09,hancock10,
jiang13}; at $T_N$ the resistivity drops discontinuously, followed by a metallic response
decreasing steadily with temperature.
Optical works on Fe$_{1+\delta}$Te observe a rapid increase in the low-frequency optical
conductivity below $T_N$ that has been attributed to a reduction of the free-carrier
scattering rate \cite{chen09,hancock10}.
Electronic structure calculations reveal the multiband nature of this material with
three hole-like bands at the center of the Brillouin zone and two electron-like bands
at the corners \cite{subedi08}; this result is in agreement with Hall-effect
measurements in Fe$_{1+\delta}$Te where both the electron and hole
carriers are observed \cite{tsukada11}.  Angle-resolved photoemission
spectroscopy (ARPES) is able to observe some of the predicted bands, but they are
strongly renormalized with a large mass enhancement $m^\ast/m_b \approx 6 - 20$,
suggesting a strongly-correlated metal \cite{xia09,tamai10,zhang10,liu13,qazilbash09}.
The enhancement of the effective mass in the PM state has been attributed to an orbital
blocking mechanism \cite{yin11}.  In addition, a pseudogap of $\simeq 65$~meV has been
observed on the electron pocket in the PM state that closes in the AFM state \cite{lin13}.
The multiband nature of this material suggests that an analysis of the optical data that
considers two different types of free-carrier contributions is in order \cite{wu10}.

%
%
In this Letter we report on the detailed temperature dependence of the in-plane optical
conductivity of Fe$_{1.03}$Te above and below $T_N$ where the electron and hole pockets
have been treated as two separate electronic subsystems.  We infer that the hole pocket
has a large scattering rate, resulting in an almost incoherent response that is largely
temperature independent.  This is in contrast to the low-frequency conductivity associated
with the electron pocket, which while roughly constant above $T_N$, increases anomalously at low
temperature.  This behavior is accompanied by a transfer of spectral weight over much of the
mid-infrared region to low energy.  The striking increase in the low-frequency conductivity
is due to a dramatic increase in the plasma frequency associated with the electron pocket
below $T_N$, suggesting that we are observing the closing  of the pseudogap with a commensurate
increase in the carrier concentration, as well as a possible reduction in the effective mass.

\begin{figure}[t]
%
%
\centerline{\includegraphics[width=3.2in]{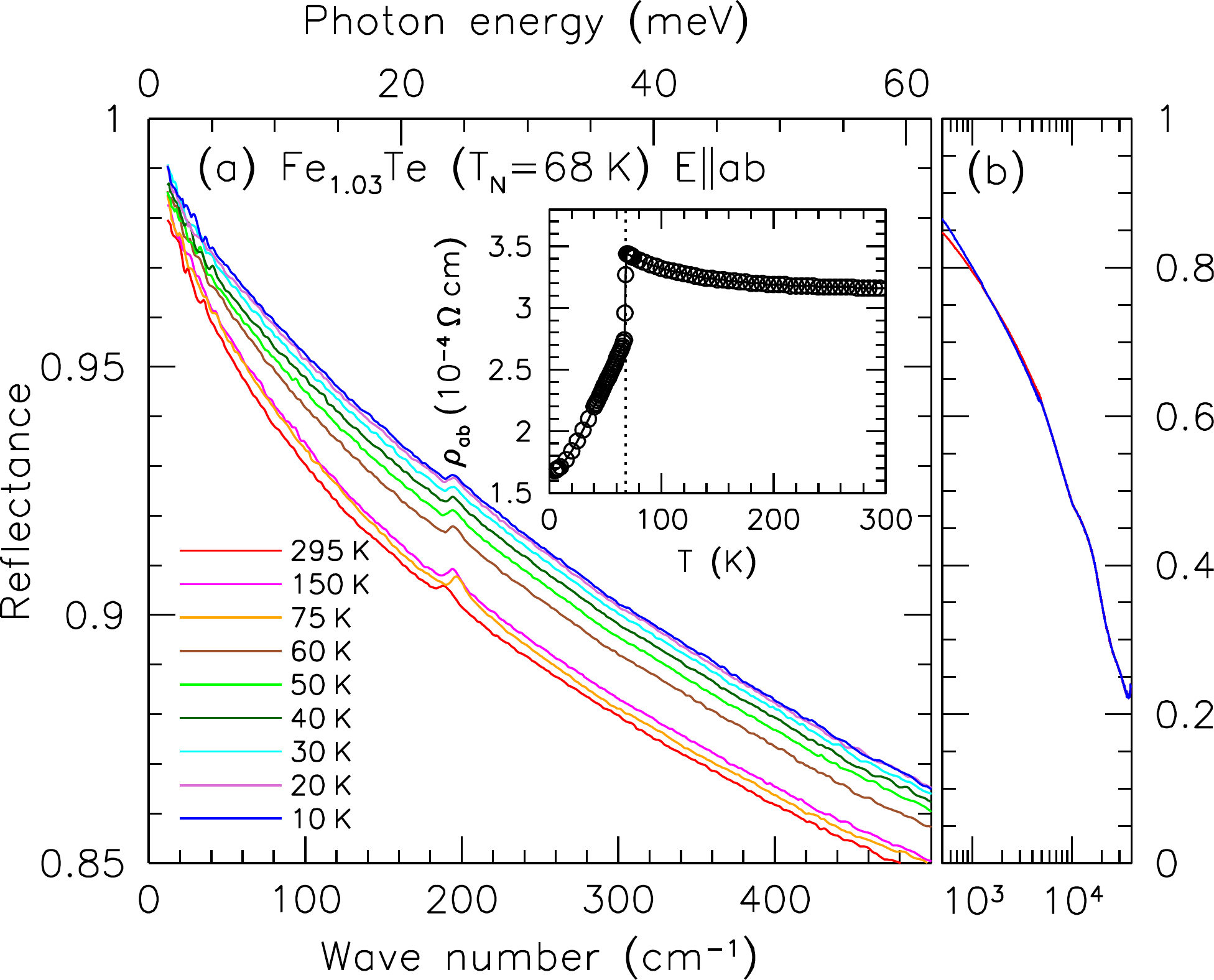}}
\caption{(a) The reflectance of Fe$_{1.03}$Te in the far-infrared region
for light polarized in the Fe-Te planes at several temperatures above and below
$T_N$.  Inset: The temperature dependence of the in-plane resistivity.  (b) The
high-frequency reflectance at 295 and 5~K.}
\label{fig:reflec}
\end{figure}

%
%
Single crystals with good cleavage planes (001) were grown by a unidirectional
solidification method with a nominal composition of Fe$_{1.03}$Te and a magnetic
and structural transition at $T_N = 68$~K.
The reflectance from the cleaved surface of a mm-sized single crystal
has been measured at a near-normal angle of incidence for several temperatures
above and below $T_N$ over a wide frequency range ($\sim 2$~meV to 4~eV) for
light polarized in the {\em a-b} planes using an {\em in situ} overcoating
technique \cite{homes93}.
The temperature dependence of the reflectance of Fe$_{1.03}$Te is shown in
Fig.~\ref{fig:reflec}(a) in the far-infrared region and at higher energies
in Fig.~\ref{fig:reflec}(b). Above $T_N$ the reflectance is metallic but
shows relatively little temperature dependence; however, below
$T_N$ the reflectance increases rapidly in this region.
This behavior is consistent with the resistivity (shown in the inset) which increases
slightly from room temperature to just above $T_N$, but which decreases rapidly below
$T_N$.  In order to understand the metallic behavior of this material below $T_N$, the
complex conductivity has been determined from a Kramers-Kronig analysis of the
reflectance \cite{dressel-book}, which requires the reflectance to be determined
over the entire frequency interval.  Given the conducting nature of this material,
the Hagen-Rubens form for the reflectance is used in the $\omega \rightarrow 0$
limit, $R(\omega) = 1-a\sqrt{\omega}$, where $a$ is chosen to match the data
at the lowest-measured frequency point.   Above the highest-measured frequency
the reflectance is assumed to be constant up to $7.5\times 10^4$~cm$^{-1}$, above
which a free electron gas asymptotic reflectance extrapolation $R(\omega)
\propto 1/\omega^4$ is assumed \cite{wooten}.

%
%

\begin{figure}[b]
%
%
\centerline{\includegraphics[width=3.0in]{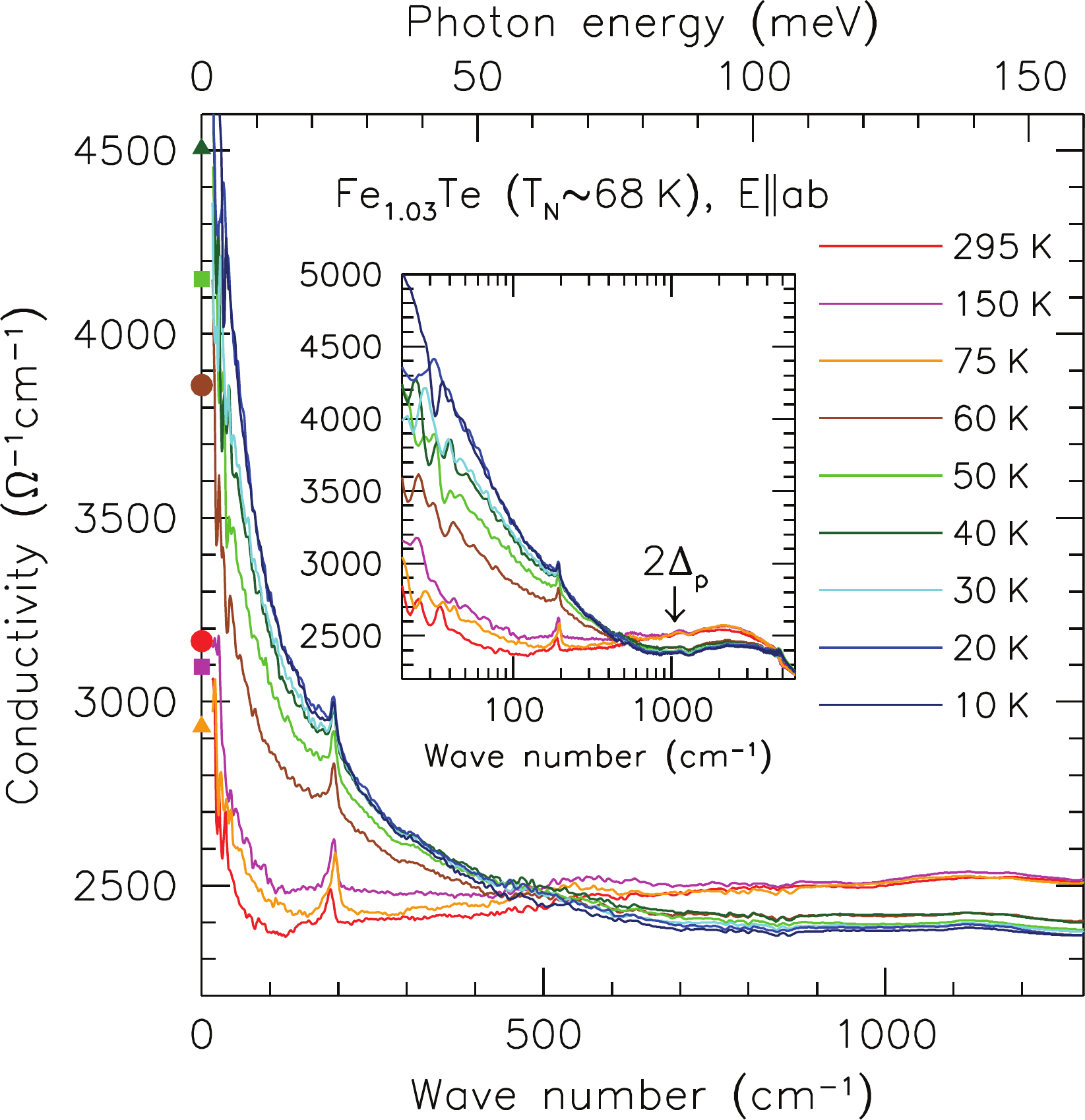}}
\caption{The temperature dependence of the real part of the optical conductivity
of Fe$_{1.03}$Te for light polarized in the \emph{a-b} planes above and below $T_N$
(the origin for the y axis is offset).  The symbols denote the values for the dc
conductivity obtained from transport measurements.
Inset: The temperature dependence of the optical conductivity shown over a
larger energy scale showing the transfer of spectral weight from high to low
energy below $T_N$.  The arrow indicates the energy scale in the optical conductivity
for the pseudogap on the electron pocket in the PM state.}
\label{fig:sigma}
\end{figure}

\begin{figure}[t]
%
\centerline{\includegraphics[width=3.0in]{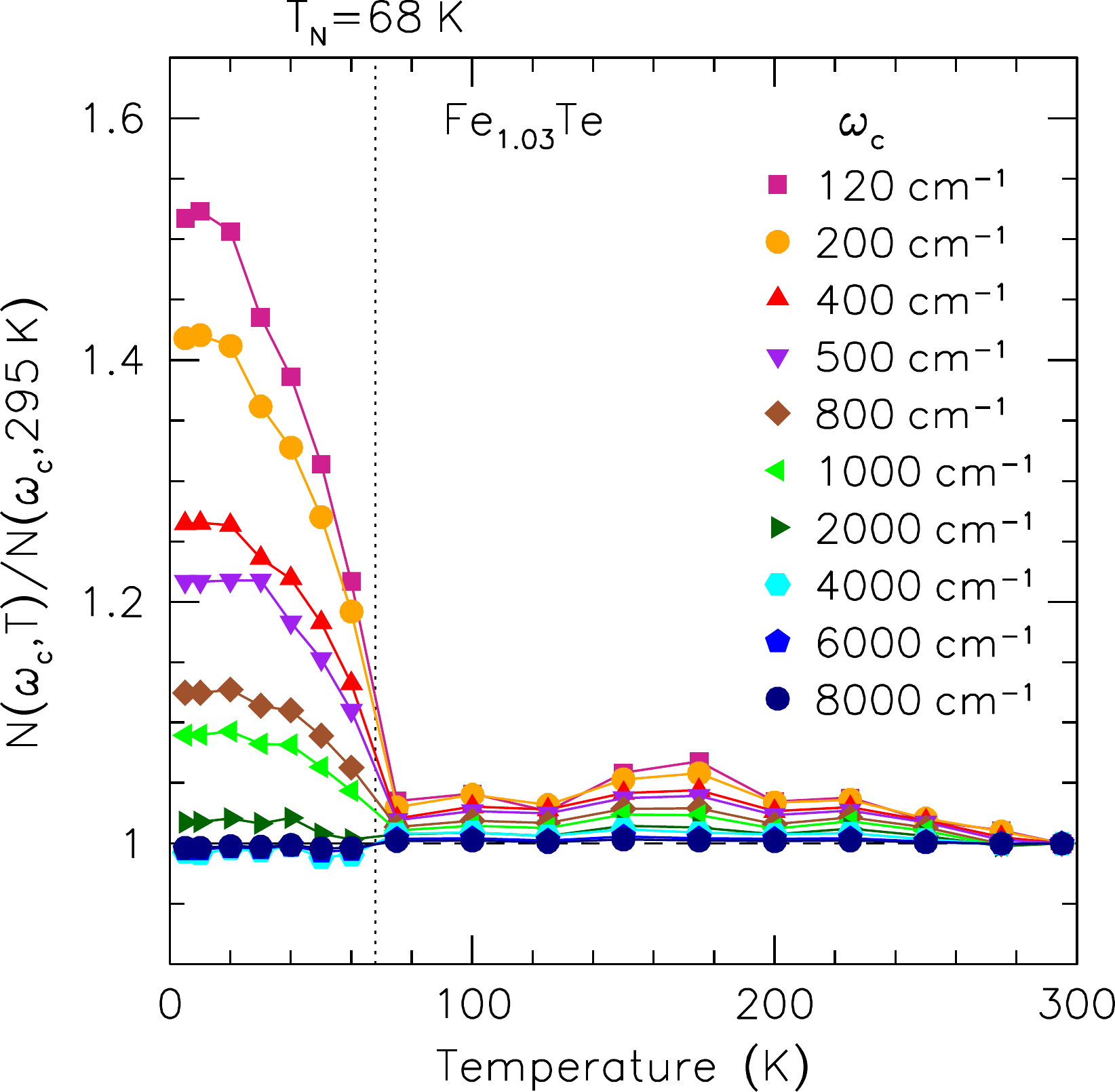}}
\caption{The temperature dependence of the normalized spectral weight
as a function of cut-off frequency, $\omega_c$. For values of $\omega_c \gtrsim
4000$~cm$^{-1}$ the spectral weight is completely captured and no
temperature dependence is observed.}
%
\label{fig:sw}
\end{figure}

%
%
The real part of the in-plane optical conductivity is shown in Fig.~\ref{fig:sigma}
for several temperatures above and below $T_N$.  The conductivity is difficult to
explain within a simple Drude model, which would follow a $\sigma_1(\omega) =
\sigma_0/(1+\omega^2\tau^2)$ form, where $\sigma_0=\sigma_1(\omega\rightarrow 0)$
and $1/\tau$ is the free-carrier scattering rate.  Above $T_N$ the conductivity is
essentially frequency and temperature independent over much of the infrared region;
only the region below $\simeq 100$~cm$^{-1}$ shows any indication of following a
Drude-like response.  Below $T_N$ there is a small but abrupt decrease in the
conductivity over much of the mid-infrared region at the same time that the
Drude-like component increases dramatically in strength, indicating a transfer of
spectral weight from high to low frequency.  The spectral weight is defined as the
area under the conductivity curve over a given interval $N(\omega_c,T)=\int_0^{\omega_c}
\sigma_1(\omega,T)\,d\omega$.
%
%
The temperature dependence of the normalized spectral weight for different
cut-off frequencies is shown in Fig.~\ref{fig:sw}.  For small cut-off frequencies
the spectral weight is roughly constant for $T>T_N$; however, for $T<T_N$ it
increase quickly.  As the cut-off frequency is increased this behavior becomes
less pronounced until by $\omega_c \simeq 4000$~cm$^{-1}$ the spectral weight
of the free carriers has been completely captured and a temperature dependence is
no longer observed.  A redistribution of spectral weight below $T_N$ has also
been observed in ARPES \cite{zhang10,lin13}.

%
%

In a single-band material, the evolution of a low-frequency Drude response to an incoherent
mid-infrared band may be explained in one of two ways.  First, the Drude response may be
superimposed on interband transitions, modeled as Lorentz oscillators, that begin at very
low frequency; while there is evidence for transitions at energies as low as 30~meV in the
iron-based materials \cite{marsik13,valenzuela13}, this is still not low enough to reproduce
incoherent component observed in this material.  A second approach is to assume a
strongly-renormalized scattering rate due to electronic correlations \cite{allen77}.
However, Fe$_{1+\delta}$Te is a multiband material \cite{subedi08}, negating the application
of the single-band model to this material.  For simplicity, the multiple hole and electron
bands are gathered into single electron and hole pockets that are treated as two separate
electronic subsystems using the so-called two-Drude model \cite{wu10} with the complex dielectric
function $\tilde\epsilon=\epsilon_1+i\epsilon_2$,
\begin{equation}
  \tilde\epsilon(\omega) = \epsilon_\infty - \sum_{j=1}^2 {{\omega_{p,D;j}^2}\over{\omega^2+i\omega/\tau_{D,j}}}
    + \sum_k {{\Omega_k^2}\over{\omega_k^2 - \omega^2 - i\omega\gamma_k}},
\end{equation}
where $\epsilon_\infty$ is the real part of the dielectric function at high frequency.
In the first sum $\omega_{p,D;j}^2 = 4\pi n_je^2/m^\ast_j$ and $1/\tau_{D,j}$ are the
square of the plasma frequency and scattering rate for the delocalized (Drude) carriers
in the $j$th band, respectively, and $n_j$ and $m^\ast_j$ are the carrier concentration
and effective mass.  In the second summation, $\omega_k$, $\gamma_k$ and $\Omega_k$ are
the position, width, and strength of the $k$th vibration or bound excitation.  The
complex conductivity is $\tilde\sigma(\omega) = \sigma_1 +i\sigma_2 = -i\omega
[\tilde\epsilon(\omega) - \epsilon_\infty ]/60$ (in units of $\Omega^{-1}$cm$^{-1}$).
%
%
%
%
The fits to the optical conductivity using this approach are in general quite good
and reveal a strong, broad Drude component in combination with a narrow, much weaker
Drude component; in addition, there are several strong, overdamped Lorentzian oscillators
in the mid- and near-infrared regions.  Below $T_N$ there is a suppression of the
optical conductivity over much of the mid-infrared region and a transfer of spectral
weight from high to low energy.  This result is in agreement with previous optical
work \cite{hancock10}, and is similar to what is observed in the normal state of
the superconducting iron-arsenic material Ba$_{0.6}$K$_{0.4}$Fe$_2$As$_2$ \cite{dai13}.
The strong Lorentzian oscillator centered in the mid infrared shifts to a somewhat
higher frequency below $T_N$, but its strength does not vary and it broadens only
slightly.  The results of the fits at several temperatures above and below $T_N$
are summarized in Table~\ref{tab:tdrude}.

%
%
We infer from the observation of the hole-like band determined from scattering
interference and the virtual absence of the electron band in a related
material \cite{allan12}, Hall effect \cite{tsukada11} and the calculated band
structure \cite{singh08}, that the broad, temperature-independent Drude component
is associated with the hole pockets, and that the narrow, strongly temperature-dependent
component is associated with the electron pockets.

\begin{figure}[b]
%
\centerline{\includegraphics[width=3.0in]{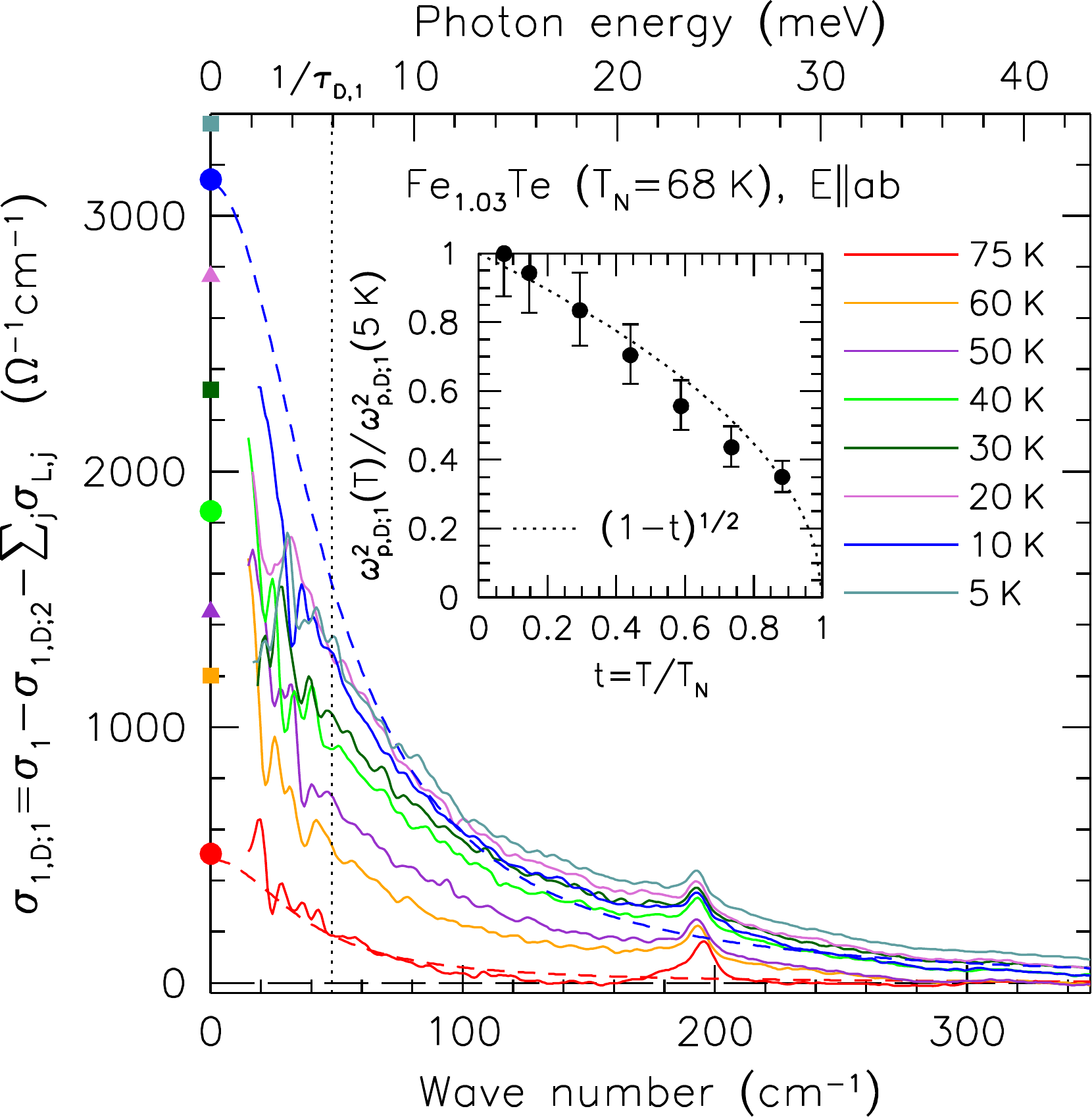}}
\caption{The low-frequency in-plane optical conductivity of Fe$_{1.03}$Te with
the broad Drude contribution and the Lorentz oscillators removed.  The symbols
at the origin denote $\sigma_{dc}-\sigma_{D2}$ (see text).  The dotted line denotes
$1/\tau_{D,1}=6$~meV, while the dashed lines are Drude fits at 75 and 10~K using
the values from Table~\ref{tab:tdrude}.
Inset: The square of the normalized plasma frequency for the narrow Drude component
versus the reduced temperature; the dotted line denotes mean-field behavior.}
%
\label{fig:remove}
\end{figure}

%
%
Although the optical conductivity is nearly frequency independent over much
of the infrared region and displays little temperature dependence after entering
the PM state, there is a dramatic increase in the low-frequency optical
conductivity below $T_N$.  To isolate this behavior, the contributions of the
broad Drude term and the Lorentzian oscillators (Table~\ref{tab:tdrude}) have
been removed, resulting in a residual component which we infer is due to the
contribution to the optical conductivity from the electron pocket,
\begin{equation}
  \sigma_{1,D;1}=\sigma_1-\sigma_{1,D;2}-\sum_j \sigma_{1,L;j}.
\end{equation}
The resulting series of curves, shown in Fig.~\ref{fig:remove}, have a
Drude-like response.
Using the contribution to the conductivity from the broad Drude term,
$\sigma_{D2}=\omega^2_{p,D;2}\tau_{D,2}/60$, and the values determined from
transport, $\sigma_{dc}=1/\rho_{ab}$, the dc conductivity values for
the electron pocket should be $\sigma_{1,D;1}(\omega \rightarrow 0) =
\sigma_{dc}-\sigma_{D2}$; these symbols are shown at the origin of
Fig.~\ref{fig:remove}.  In the zero-frequency limit the conductivity
extrapolates to these values reasonably well.
%
%
%
\begin{table}[t]
\caption{The temperature dependence of the Drude parameters for the electron
and hole pockets and the Lorentz oscillators (bound excitations); the estimated
uncertainties are for the fitted parameters is approximately 5\%.
All units are in cm$^{-1}$ unless otherwise indicated.}
\begin{ruledtabular}
\begin{tabular}{c cc c cc}

  T (K) &
  $\omega_{p,D;1}$ & $1/\tau_{D,1}$ & $1/\tau^{a}_{D,2}$ & $\omega^{b}_{L,1}$ & $\gamma^{b}_{L,1}$ \\
\cline{1-6}
%
%
   295 &  1010 & $\sim 28$ & 805 & 3030 & $12\,110$ \\
   150 &  1200 & $\sim 38$ & 776 & 3004 & $12\,016$ \\
   100 &  1180 & $\sim 40$ & 780 & 3027 & $12\,001$ \\
    75 &  1050 & $\sim 38$ & 786 & 3023 & $12\,022$ \\
    60 &  1850 & $\sim 48$ & 788 & 3285 & $12\,391$ \\
    50 &  2035 & $\sim 48$ & 765 & 3305 & $12\,470$ \\
    40 &  2300 & $\sim 48$ & 774 & 3278 & $12\,399$ \\
    30 &  2570 & $\sim 48$ & 784 & 3335 & $12\,462$ \\
    20 &  2820 & $\sim 48$ & 782 & 3342 & $12\,473$ \\
    10 &  3010 & $\sim 48$ & 772 & 3350 & $12\,505$ \\
\end{tabular}
\end{ruledtabular}
\footnotetext[1] {$T>T_N$, $\omega_{p,D;2}=10\,700$;
  $T<T_N$, $\omega_{p,D;2}= 11\,200$~cm$^{-1}$.}
\footnotetext[2] {$\Omega_{L,1} = 40\,800$~cm$^{-1}$.}
\label{tab:tdrude}
%
%
\end{table}
%
The Drude fits to the narrow component have been constrained to agree with
these values and indicate that while $\omega_{p,D;1}$ is increasing rapidly
below $T_N$, the scattering rate $1/\tau_{D,1}$ is roughly constant.
In fact, given that the Drude profile is a Lorentzian centered at zero
frequency and the width at half maximum is the scattering rate, the fact
that $1/\tau_{D,1}$ is not changing may be observed simply by inspection
(dotted line in Fig.~\ref{fig:remove}).  The two dashed lines in
Fig.~\ref{fig:remove} show the results of the fits just above $T_N$ and
for $T\ll T_N$; in general the fits close to $T_N$ are quite good while
for $T\ll T_N$ the fits are somewhat poorer, perhaps because there
are in fact two electron pockets rather than just one we have used to
model the low-frequency conductivity.

%
%
The rapid increase of the low-frequency conductivity below $T_N$ is striking.
Normally, this behavior is associated with a decrease in the scattering rate,
resulting in a narrowing of the Drude feature.  While this narrowing results
in the transfer of spectral weight from high to low frequency, the total
weight associated with the individual Drude component should remain constant.
However, not only does the scattering rate actually increase slightly
upon entering the AFM state (whereupon it remains roughly constant
below $T_N$), the spectral weight is observed to increase dramatically,
indicating that is in fact the plasma frequency that is increasing.
The temperature dependence of $\omega^2_{p,D;1}$ below $T_N$ shown
in the inset of Fig.~\ref{fig:remove} follows a mean-field temperature
dependence.
Given the definition of $\omega^2_{p,D;1}=4\pi n_1 e^2/m_1^\ast$, then either
the carrier concentration is increasing or the effective mass is decreasing.
However, in this instance there is evidence to suggest that both quantities are
changing below $T_N$.

%
%

In the PM state there is a partial gap (pseudogap) on the electron pocket of
$\Delta_p \simeq 65$~meV (525~cm$^{-1}$) \cite{lin13}.  In the optical conductivity
a pseudogap in the spectrum of excitations shifts spectral weight associated with
the free-carriers in the electron pocket to energies above $2\Delta_p \simeq 130$~meV
(1050~cm$^{-1}$) resulting in a localized, incoherent response.  This is precisely what
is observed in Fig.~\ref{fig:sw}, where $\omega_c > 2\Delta_p$ is necessary to recover
the full spectral weight.  However, below $T_N$ the pseudogap is observed to close \cite{lin13}
resulting in coherent charge transport; this is consistent with the appearance of quasiparticles
below $T_N$ \cite{zhang10}. As a result, the number of carriers in the electron pocket ($n_1$)
will increase below $T_N$ with a commensurate increase in the plasma frequency,
$(\omega_{p,D;1})$.
In addition, there is also evidence that in the PM state of FeTe there is a large mass
enhancement of the Fe $d_{xy}$ orbital of $m^\ast/m_{band} \simeq 7$; this effect is
attributed to an orbital-blocking mechanism \cite{yin11}.  In the AFM state the spin
fluctuations are observed to decrease \cite{zhang10} and most signs of correlations
disappear \cite{lin13}, suggesting that orbital blocking, and as a consequence the
effective mass, may be reduced below $T_N$.  Thus, the increase in $n_1$ and the
decrease in $m_1^\ast$ may combine in tandem to lead to an increase of the plasma
frequency for the carriers on the  electron pocket in the AFM state.

%
%
To conclude, we have measured the in-plane optical properties of Fe$_{1.03}$Te above and
below $T_N$.  The optical conductivity has been described using the two-Drude model,
revealing a strong component with a large scattering rate associated with the hole
pocket that varies little with temperature, and a weaker component with a much smaller
scattering rate associated with the electron pocket that displays a strong temperature
dependence below $T_N$.  The plasma frequency associated with the carriers in the electron
pocket increases anomalously below $T_N$ due to an increase in the carrier concentration
following the closing of the pseudogap, as well as the possible decrease of the effective mass
in the AFM state.

%
%
The authors thank J. C. Davis, K. Haule, J. Hwang, G.~Kotliar, J. H. Shim,
and I.~Zaliznyak for helpful discussions.
Research supported by the U.S. Department of Energy (DOE), Office of
Basic Energy Sciences, Division of Materials Sciences and Engineering
under Contract No. DE-AC02-98CH10886.  RDZ and JS were supported by the
Center for Emergent Superconductivity, an Energy Frontier Research
Center, DOE.

%
%
%
%

%

\end{document}